\documentclass
[superscriptaddress,secnumarabic,amssymb,amsmath,nobibnotes,aps,prd,showkeys,showpacs,nofootinbib,onecolumn]{revtex4}%
\usepackage{graphicx}
\usepackage{epsf}
\usepackage{bm}
\usepackage{amsmath}
\usepackage{amsfonts}
\usepackage{amssymb}
\usepackage{xcolor}
\usepackage{epstopdf}%
\providecommand{\U}[1]{\protect\rule{.1in}{.1in}}
\newcommand{\be}{\begin{equation}}
\newcommand{\ee}{\end{equation}}

\newcommand{\mincir}{\raise
-3.truept\hbox{\rlap{\hbox{$\sim$}}\raise4.truept\hbox{$<$}\ }}
\newcommand{\magcir}{\raise
-3.truept\hbox{\rlap{\hbox{$\sim$}}\raise4.truept\hbox{$>$}\ }}

\begin{document}
\title{Integrability of the Einstein-nonlinear $SU(2)$ $\sigma$-model in a nontrivial
topological sector}
\author{Andronikos Paliathanasis}
\email{anpaliat@phys.uoa.gr}
\affiliation{Instituto de Ciencias F\'{\i}sicas y Matem\'{a}ticas, Universidad Austral de
Chile, Valdivia, Chile}
\affiliation{Institute of Systems Science, Durban University of Technology, PO Box 1334,
Durban 4000, Republic of South Africa}
\author{Tim Taves}
\email{timtaves@gmail.com}
\affiliation{Centro de Estudios Cient\'{\i}ficos (CECS), Arturo Prat 514, Valdivia, Chile}
\author{P.G.L. Leach}
\email{leach.peter@ucy.ac.cy}
\affiliation{Department of Mathematics and Institute of Systems Science, Research and
Postgraduate Support, Durban University of Technology, PO Box 1334, Durban
4000, Republic of South Africa}
\affiliation{School of Mathematics, Statistics and Computer Science, University of
KwaZulu-Natal, Private Bag X54001, Durban 4000, Republic of South Africa}

\begin{abstract}
The integrability of the\ $\Lambda-$Einstein-nonlinear $SU(2)$ $\sigma$-model
with nonvanishing cosmological charge is studied. We apply the method of
singularity analysis of differential equations and we show that the equations
for the gravitational field are integrable. The first few terms of the
solution are presented.

\end{abstract}
\keywords{$\sigma$-model; Pions; Integrability; Singularity analysis}
\pacs{98.80.-k, 95.35.+d, 95.36.+x}
\maketitle
\date{\today}

\noaffiliation

%

%TCIMACRO{\TeXButton{bigskip}{\bigskip}}%
%BeginExpansion
\bigskip
%EndExpansion

Nonlinear $\sigma-$models are important theoretical models in Physics for the
properties that they provide which are of special interest \cite{ref2,ref3}.
The Einstein nonlinear $\sigma-$models, in which the total Action Integral is
the sum of the Einstein-Hilbert Action Integral and the Action Integral which
corresponds to the nonlinear matter source, have provided different kinds of
solutions for the gravitational equations. Specifically it has been shown that
there exist black hole solutions with a regular event horizon which
asymptotically approach the Schwarzschild spacetime, in the context of the
Einstein-Skyrme Model, which violates the \textquotedblleft no
hair\textquotedblright\ conjecture for black holes (see for instance
\cite{Ioannidou} and references therein).

The purpose of this work is to study the integrability of the field equations
of the Einstein-nonlinear~$SU\left(  2\right)  $ $\sigma-$model, which have
been studied previously in \cite{canzan}. We do this by using the method of
singularity analysis of differential equations\footnote{For a review on the
singularity analysis see \cite{buntis} and subsequent developments in
\cite{Feix97a,Andriopoulos06a}}. The application of singularity analysis in
gravitational theories has been applied by many researchers in the past, for
instance in the case of the Mixmaster Universe (Bianchi IX)
\cite{CotsakisLeach,Con1,Demaret}, in scalar field cosmology \cite{miritzis}
and in modified theories of gravity \cite{aleach,arn,arn2}.

Consider a Riemannian manifold $\mathit{M}$ with metric $g_{\mu\nu}$ of
Lorentzian signature. The action integral of the field equations for the
Einstein-nonlinear $SU\left(  2\right)  $ $\sigma-$model in a four-dimensional
manifold is given by%
\begin{equation}
S=S_{EH}+S_{\left(  \sigma\right)  }, \label{eq.00}%
\end{equation}
where $S_{EH}$ is the Einstein-Hilbert action with the cosmological constant,
i.e., $S_{EH}=\int dx^{4}\left(  R-2\Lambda\right)  $, and $S_{\left(
\sigma\right)  }$ is the action integral of the nonlinear sigma
model~\cite{Nair}%
\begin{equation}
S_{\left(  \sigma\right)  }=\frac{K}{2}\int dx^{4}\sqrt{-g}\left(  \left(
U^{-1}U_{;\mu}\right)  g^{\mu\nu}\left(  U^{-1}U_{;\nu}\right)  \right)  ,
\label{eq.01}%
\end{equation}
where $U\left(  x^{\nu}\right)  $ is the $SU\left(  2\right)  $-valued scalar
and $K$ is a positive constant. The physical implication of the action,
(\ref{eq.01}), is that it describes the dynamics of low energy pions. The
gravitational field equations are derived by variation of the action integral
(\ref{eq.00}) with respect to the metric tensor $g_{\mu\nu}$. This leads to
the following set of equations,
\begin{equation}
G_{\mu\nu}+\Lambda g_{\mu\nu}=T_{\mu\nu}, \label{eq.02}%
\end{equation}
in which the left hand side of (\ref{eq.02}) corresponds to the
Einstein-Hilbert action where $G_{\mu\nu}$ is the Einstein tensor and
$\Lambda$ is the cosmological constant. The right hand side of (\ref{eq.02})
is that of the nonlinear $\sigma$-model and provides the matter source. The
explicit form of the energy-momentum tensor is
%\begin{align}
%T_{\mu\nu}  &  =-\frac{K}{2}\left(  U^{-1}U_{;\mu}\right)  \left(
%U^{-1}U_{;\nu}\right)  \label{eq.03}\\
%&  ~~\frac{K}{4}g_{\mu\nu}\left(  U^{-1}U_{;\kappa}\right)  g^{\kappa\lambda
%}\left(  U^{-1}U_{;\lambda}\right).
%\end{align}%
\begin{equation}
T_{\mu\nu}=-\frac{K}{2}\left(  U^{-1}U_{;\mu}\right)  \left(  U^{-1}U_{;\nu
}\right)  +\frac{K}{4}g_{\mu\nu}\left(  U^{-1}U_{;\kappa}\right)
g^{\kappa\lambda}\left(  U^{-1}U_{;\lambda}\right)  . \label{eq.03}%
\end{equation}
Furthermore variation with respect to the scalar-valued $U\left(  x^{\nu
}\right)  $, in (\ref{eq.00}) leads to the constraint equation $\left(
U^{-1}U_{;\mu}\right)  _{;\nu}g^{\mu\nu}=0,~$while the latter can follow from
the application of the Bianchi identity in (\ref{eq.02}), that is, $T_{~~;\nu
}^{\mu\nu}=0$.

By following the Ansatz, which was proposed in \cite{can01,can02,can02a,can03}
and its generalizations \cite{can4,can5,can6,can7}, for the parametrization of
the $SU\left(  2\right)  $ algebra in \cite{canzan} Ay\'{o}n-Beato, Canfora
and Zanelli found that for the four-dimensional spacetime,
\begin{equation}
ds^{2}=-F(r)\left(  dt+\cos\theta d\varphi\right)  ^{2}+N(r)^{2}dr^{2}%
+\rho^{2}(r)\left(  d\theta^{2}+\sin^{2}\theta d\varphi^{2}\right)  ,
\label{eq.04a}%
\end{equation}
the conservation equation, $T_{~;\nu}^{\mu\nu}=0,~$is satisfied always and the
energy-momentum tensor, (\ref{eq.03}), is expressed only in terms of the
fields $F$, $\rho$ and $N$. Specifically, the matter field equations $\left(
U^{-1}U_{;\mu}\right)  _{;\nu}g^{\mu\nu}=0$, are identically satisfied so that
one has only to deal with the Einstein's field equations while the solution
provides a nontrivial topological sector \cite{canzan}.

Spacetime, (\ref{eq.04a}), is a locally rotational spacetime and for arbitrary
functions, $F$ and $\rho$, admits a four-dimensional Killing Algebra which
comprises the autonomous symmetry, $A_{1}=\left\{  \partial_{t}\right\}  $,
and $SO\left(  3\right)  $, i.e., the Killing vectors form the $A_{1}\oplus
SO\left(  3\right)  $ Lie Algebra \cite{tsaLRS}.

In the minisuperspace approach the gravitational field equations (\ref{eq.02})
can arise from the Euler-Lagrange equations of the singular Lagrangian,%

\begin{align}
L\left(  F,F^{\prime},\rho,\rho^{\prime},N\right)   &  =\frac{4}{N}\left(
\sqrt{F}F^{\prime2}+\frac{\rho}{\sqrt{F}}F^{\prime}\rho^{\prime}\right)
+\Lambda\rho^{2}N\sqrt{F}\nonumber\\
&  \quad-\frac{K}{2}NF^{-\frac{1}{2}}\left(  \rho^{2}-2F\right)  -\rho
^{-2}N\sqrt{F}\left(  F+4\rho^{2}\right)  , \label{eq.04b}%
\end{align}
where the equation
\begin{equation}
\frac{\partial L}{\partial N}=0
\end{equation}
gives the constraint equation or the
\begin{equation}
G_{r}^{r}-T_{r}^{r}=0
\end{equation}
component of (\ref{eq.02}). In (\ref{eq.04b}) we can see the dynamical terms
which corresponds to the $R^{\left(  3\right)  }$ curvature term of
(\ref{eq.04a}) of the cosmological constant and of the $\sigma-$model.

Because the field equations are singular there could exist a nonlocal
conservation law which is generated by the conformal Killing vectors of the
minisuperspace, for details see \cite{dim1,dim2}. In our consideration, as the
minisuperspace of (\ref{eq.04b}) has dimension two, there exists an infinite
number of conformal Killing vectors and an infinite number of nonlocal
conservation laws. Hence with the use of a nonlocal conservation law the two
second-order differential equations and the first-order differential equation,
which describe the gravitational field equations, can be reduced to the
second-order nonautonomous equation \cite{CanZ},%
\begin{align}
0  &  =ry\left(  Kr^{6}-2r^{3}\left(  K-4+4r^{2}\Lambda\right)  y+2y^{2}%
\right)  y^{\prime\prime}\nonumber\\
&  \quad-\left(  6Kr^{6}\left(  y^{2}\right)  ^{\prime}+Kr^{7}\left(
y^{\prime}\right)  ^{2}-\frac{3}{2}\left(  y^{4}\right)  ^{\prime}-\frac{2}%
{3}r\left(  y^{3}\right)  ^{\prime}\left(  8r^{4}\Lambda+y^{\prime}\right)
\right)  , \label{eq.05}%
\end{align}
where $y=y\left(  r\right)  =\rho F,$ $N\left(  r\right)  $ is
\begin{equation}
N\left(  r\right)  =\frac{\left(  2y^{2}-Kr^{6}\right)  }{2y^{2}\left(
4r^{3}+y-4\Lambda r^{5}\right)  +Ky\left(  r^{6}-2r^{3}y\right)  }%
\frac{h^{\prime}\left(  r\right)  }{V_{eff}}, \label{eq.06}%
\end{equation}
in which
\begin{align}
\left(  r^{3}\left(  y^{2}\right)  ^{\prime}\right)  \left(  h(r)\right)
^{2}  &  =\left(  2K\,r^{6}-16\,\Lambda r^{5}\right) \label{eq.07}\\
&  \quad+4\,yr^{3}\left(  \left(  4-K\right)  +4y\right)  , \label{eq.09}%
\end{align}
and the new gauge has been selected to be so that $\rho=r$ . The term
$V_{eff}$ in (\ref{eq.06}) includes all the potential terms of (\ref{eq.04b})
so that the Lagrangian (\ref{eq.04b}) is that of geodesic equations in a
two-dimensional manifold.

The nonautonomous equation (\ref{eq.05}) can always be written in the form of
an autonomous third-order differential equation by introducing the new
variables $r\rightarrow Y\left(  x\right)  $ and $y\rightarrow Y_{,x}$. The
third-order equation is%
\begin{align}
0  &  =Y^{6}Y_{x}\left(  8\Lambda\left(  Y_{,xx}\right)  ^{2}-Y_{,x}\left(
3KY_{,xx}+8\Lambda Y_{,xxx}\right)  \right) \nonumber\\
&  \quad+Y^{7}\left(  KY_{,x}Y_{,xxx}-2K\left(  Y_{,xx}\right)  ^{2}\right)
+2(K-4)Y^{4}Y_{,x}\left(  \left(  Y_{,xx}\right)  ^{2}-Y_{,x}Y_{,xxx}\right)
\nonumber\\
&  \quad+2Y\left(  Y_{,x}\right)  ^{3}Y_{,xxx}+\left(  16\Lambda
Y^{5}+6\left(  Y_{,x}\right)  \right)  \left(  Y_{,x}\right)  ^{3}Y_{,xx},
\label{eq.10}%
\end{align}
for which reduction with the autonomous symmetry, $\partial_{x}$, leads to the
original equation, (\ref{eq.05}).

We found that (\ref{eq.10}), except the autonomous one, does not admit any
other point symmetry vector for any value of the cosmological constant. That
is an interesting result because it indicates that there exists a unique
relation among equations (\ref{eq.05}) and (\ref{eq.10}). On the other hand,
it has been found in \cite{CanZ} that equation (\ref{eq.05}) admits a
rescaling symmetry when the cosmological constant is zero. The application of
this symmetry vector reduced equation (\ref{eq.05}) to a first-order Abel
equation. However, for nonvanishing cosmological constant only special
solutions of the form $y\left(  r\right)  =\sigma_{0}r^{3}$ have been derived,
where $\sigma_{0}=-1,\frac{K}{4}$. Those special solutions correspond to
specific initial conditions which describe the asymptotic behaviour of the
general evolution of the system.

Below we assume the case of nonvanishing cosmological constant and we perform
the singularity analysis. Note that, if (\ref{eq.10}) is integrable, then
(\ref{eq.05}) is also integrable which means that the gravitational field
equations for the Einstein-nonlinear $SU\left(  2\right)  $ $\sigma-$model are
also integrable, that is, the dynamical system which follows from the action
integral, (\ref{eq.00}), is integrable.

We define the new variable $\Phi\left(  x\right)  =Y\left(  x\right)  ^{2}$
and we search for power-law solutions $\Phi\left(  x\right)  =\alpha\chi^{p}$
in (\ref{eq.10}) from where we have the following possible sets $(\chi
=x-x_{0})$, where $x_{0}$ is the location of the putative movable singularity)%
\begin{equation}
p=-1~\text{with}~\alpha=-\frac{2}{K}~,~\alpha=\frac{1}{2} \label{eq.11}%
\end{equation}
and%
\begin{equation}
p=-\frac{1}{2}~\text{with}~\alpha=\pm\frac{i}{2}\sqrt{\frac{3}{2\Lambda}%
}\text{.} \label{eq.12}%
\end{equation}

We consider the values of (\ref{eq.11}) from which we can see that these are
the power-law solutions of (\ref{eq.05}) which we described above. The next
step, in order to test if (\ref{eq.10}) passes the singularity test, is to
determine the resonances. Let $\alpha=-\frac{2}{K}$, which is the case in
which the solution leads to a Lorentzian signature spacetime, for details see
\cite{CanZ}. Then by substituting
\begin{equation}
\Phi\left(  \chi\right)  =\alpha\chi^{-1}+\gamma\chi^{-1+s} \label{eq.13}%
\end{equation}
into (\ref{eq.10}) and taking the terms linear in $\gamma$, we have $s\left(
2s^{2}-s-3\right)  =0$, which gives the triple solution%
\begin{equation}
s_{0}=-1~,~s_{1}=0~,~s_{2}=\frac{3}{2}. \label{eq.14}%
\end{equation}
Here we remark that the resonances are the same and for $\alpha=\frac{1}{2}$,
from where we can say that (\ref{eq.10}) passes the singularity test and it is integrable.

As far as the solution is concerned, we can write it in a series form in which
from $s_{2}$, we have that the powers of $\chi$ in the series increase by
$\frac{1}{2}$. Therefore the solution is
\begin{equation}
\Phi\left(  \chi\right)  =m_{0}\chi^{-1}+m_{1}\chi^{-\frac{1}{2}}+m_{2}%
+m_{3}\chi^{\frac{1}{2}}+%
%TCIMACRO{\dsum \limits_{I=+4}^{+\infty}}%
%BeginExpansion
{\displaystyle\sum\limits_{I=+4}^{+\infty}}
%EndExpansion
m_{I}\chi^{-1+\frac{m_{I}}{2}}, \label{eq.15}%
\end{equation}
where $m_{3}$ and $m_{0}$ are arbitrary constants\footnote{Recall that there
exists a resonance with value zero.} and $m_{1},m_{2},m_{I}$ have to be
determined. In particular they are functions of $m_{0},m_{3},K~$and $\Lambda.$

Hence we substitute (\ref{eq.15}) into (\ref{eq.10}) and find for the
coefficient constants that $m_{1}=0$,
\begin{equation}
m_{2}=\frac{3}{8m_{0}\Lambda}\left(  m_{0}\left(  4+K\left(  2m_{0}-1\right)
\right)  -2\right)  \label{eq.16}%
\end{equation}
and%

\begin{equation}
m_{4}=-\frac{3\left(  2m_{0}-1\right)  }{64m_{0}^{3}\Lambda^{2}}\left(
12+8\left(  K-1\right)  m_{0}+K^{2}m_{0}^{2}\left(  1+m_{0}\right)  \right)
\label{eq.17}%
\end{equation}
from which we can see that for $m_{0}=-\frac{2}{K}$ it follows that $m_{2}%
=0$and $m_{4}=0$. For values of $\chi$ such that $\chi^{-1}>>\chi^{2}$ the
solution takes the following form%
\begin{equation}
\Phi\left(  \chi\right)  \simeq-\frac{2}{K}\chi^{-1}+m_{3}\chi^{\frac{1}{2}}.
\label{eq.17a}%
\end{equation}

Now the spacetime, (\ref{eq.04a}), has the following form%
\begin{equation}
ds^{2}=-\frac{\Phi\left(  \chi\right)  _{,\chi}}{2\Phi\left(  \chi\right)
}\left(  dt+\cos\theta d\varphi\right)  ^{2}+\frac{N(\Phi,\Phi_{,\chi}%
)^{2}\Phi\left(  \chi\right)  _{,\chi}}{2\sqrt{\Phi\left(  \chi\right)  }%
}d\chi^{2}+\Phi\left(  \chi\right)  \left(  d\theta^{2}+\sin^{2}\theta
d\varphi^{2}\right)  . \label{eq.18}%
\end{equation}

Furthermore from the second dominant behavior $p=-\frac{1}{4}$ we find the
resonances
\begin{equation}
s_{0}=-1\text{~},~s_{1}=-\frac{1}{2}~,~s_{2}=\frac{3}{2} \label{eq.19}%
\end{equation}
which provides us with the second solution%
\begin{equation}
\Phi\left(  \chi\right)  =%
%TCIMACRO{\dsum \limits_{I=-4}^{-\infty}}%
%BeginExpansion
{\displaystyle\sum\limits_{I=-4}^{-\infty}}
%EndExpansion
n_{J}\chi^{-\frac{1+J}{2}}+n_{-2}\chi^{-\frac{3}{2}}+n_{-1}x^{-1}%
+n_{0}x^{-\frac{1}{2}}+n_{1}+n_{2}^{\frac{1}{2}}+n_{3}x+%
%TCIMACRO{\dsum \limits_{I=+4}^{+\infty}}%
%BeginExpansion
{\displaystyle\sum\limits_{I=+4}^{+\infty}}
%EndExpansion
n_{I}\chi^{-\frac{1+I}{2}} \label{eq.20}%
\end{equation}
which is a right and left Laurent expansion. The free parameters of solution
(\ref{eq.20}) are the $n_{-1}$, and $n_{4}$.

One important issue of the general solution of (\ref{eq.18}) is the number of
constants which are two from the solution (\ref{eq.15}); the $\left\{
m_{0},m_{3}\right\}  ,~\left\{  n_{-1},n_{4}\right\}  $ and the constant $K$
which corresponds to the matter source. One would expect three integration
constants for the system (\ref{eq.02}). The latter is hidden in $\chi$, as for
the singularity analysis we move to the complex plane in which $\chi=x-x_{0}$
and $x_{0}$ is the position of the singularity. However, that is not an
essential constant because it can always be absorbed with the transformation
of the coordinate, $x,$ which means that at the end there are only two constants.

\begin{figure}[ptb]
\includegraphics[height=7cm]{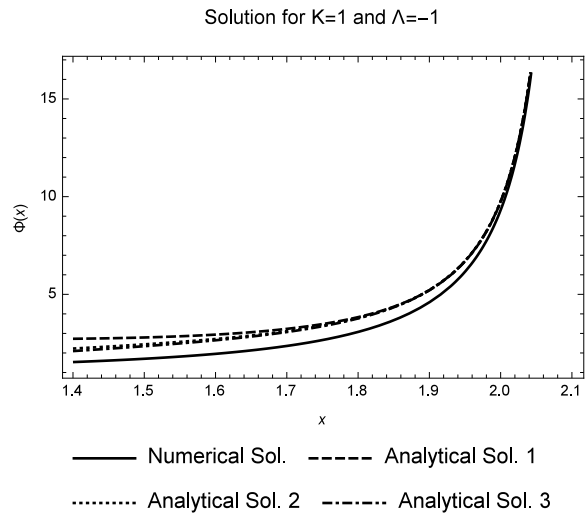}
%\mbox{\epsfxsize=14.2cm \epsffile{numericmodel2q.eps}}
\caption{Numerical simulation and analytical approximations close to the
singulartity for the master equation. }%
\label{fig1}%
\end{figure}

\begin{figure}[ptb]
\includegraphics[height=7cm]{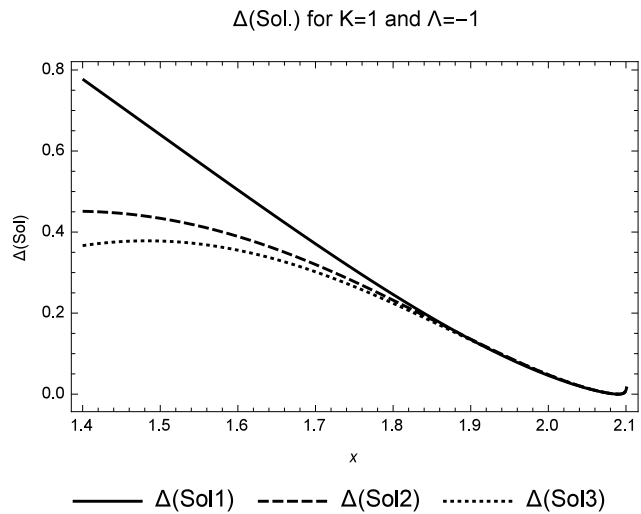}
%\mbox{\epsfxsize=14.2cm \epsffile{numericmodel2q.eps}}
\caption{Relative error of the analytical approximations with numerical
simulation for the master equation close to the singulartity. }%
\label{fig2}%
\end{figure}

On the other hand, as we discussed above, the asymptotic behaviour of the
general solution (\ref{eq.15}) is $\Phi\left(  \chi\right)  $ $\simeq\chi
^{-1}$, which is the dominant term around the singularity of (\ref{eq.10}). If
we start far from that solution, that is, from different initial conditions,
then in the asymptotic behaviour of the solution only the terms of lower
powers of $\chi$ contribute in the solution. Hence the solution can be
described well from the first terms of the Laurent series. We demonstrate that
in fig. \ref{fig1} where the numerical solution of (\ref{eq.10}) (using
Mathematica's NDSolve routine) is given and compared with analytical solutions
from the Laurent expansion (\ref{eq.15}) where we considered the first four
terms of (\ref{eq.15}), $\left(  m_{0}-m_{3}\right)  ~$(Analytic Sol.1), the six first
terms $\left(  m_{0}-m_{5}\right)  $ (Analytic Sol..2) and the seven first terms
$\left(  m_{0}-m_{6}\right)  ~$(Analytic Sol..3). Figure \ref{fig2} presents the
absolute error between the analytical solutions and the numerical solution. We
observe that those specific initial conditions, where the singularity $x_{0}$
is at $x_{0}\simeq2.10,$ \ are very close in the region of the singularity the
approximation works well. Of course eventually the error becomes big and other
terms of the Laurent expansion have to be considered.

We remark that in the case of vanishing cosmological constant the singularity
analysis provides that the resonances depend on the parameter $K$. Last but
not least an important observation which we can extract from the singularity
analysis about the stability of the leading order term is that the solution is
unstable. The reason for that is the existence of the terms given by the right
Laurent expansion which dominate as far as we are moving from the singularity,
see also \cite{sin3} and references therein.

\begin{acknowledgments}
This work has been funded by the FONDECYT grants no. 3160121 (AP) and 3140123 (TT). The Centro de Estudios Científicos (CECs) is funded by the Chilean Government through the Centers of Excellence Base Financing Program of Conicyt. PGL Leach
thanks the Instituto de Ciencias F\'{\i}sicas y Matem\'{a}ticas of the UACh
for the hospitality provided while this work carried out and acknowledges the
National Research Foundation of South Africa and the University of
KwaZulu-Natal for financial support. 
\end{acknowledgments}

\end{document}